\documentclass[10pt,a4paper,onecolumn]{article}
\usepackage{marginnote}
\usepackage{graphicx}
\usepackage{xcolor}
\usepackage{authblk,etoolbox}
\usepackage{titlesec}
\usepackage{calc}
\usepackage{tikz}
\usepackage{hyperref}
\hypersetup{colorlinks,breaklinks=true,
            urlcolor=[rgb]{0.0, 0.5, 1.0},
            linkcolor=[rgb]{0.0, 0.5, 1.0}}
\usepackage{caption}
\usepackage{tcolorbox}
\usepackage{amssymb,amsmath}
\usepackage{ifxetex,ifluatex}
\usepackage{seqsplit}
\usepackage{xstring}

\usepackage{float}
\let\origfigure\figure
\let\endorigfigure\endfigure
\renewenvironment{figure}[1][2] {
    \expandafter\origfigure\expandafter[H]
} {
    \endorigfigure
}

\usepackage{fixltx2e} 
\usepackage[
  backend=biber,
]{biblatex}
\bibliography{paper.bib}

\let\textttOrig=\texttt
\def\texttt#1{\expandafter\textttOrig{\seqsplit{#1}}}
\renewcommand{\seqinsert}{\ifmmode
  \allowbreak
  \else\penalty6000\hspace{0pt plus 0.02em}\fi}

\makeatletter
\let\href@Orig=\href
\def\href@Urllike#1#2{\href@Orig{#1}{\begingroup
    \def\Url@String{#2}\Url@FormatString
    \endgroup}}
\def\href@Notdoi#1#2{\def\tempa{#1}\def\tempb{#2}%
  \ifx\tempa\tempb\relax\href@Urllike{#1}{#2}\else
  \href@Orig{#1}{#2}\fi}
\def\href#1#2{%
  \IfBeginWith{#1}{https://doi.org}%
  {\href@Urllike{#1}{#2}}{\href@Notdoi{#1}{#2}}}
\makeatother

\newlength{\cslhangindent}
\newlength{\csllabelwidth}
\newenvironment{CSLReferences}[3] 
 {
  \setlength{\parindent}{0pt}
  \ifodd #1 \everypar{\setlength{\hangindent}{\cslhangindent}}\ignorespaces\fi
  \ifnum #2 > 0
  \setlength{\parskip}{#2\baselineskip}
  \fi
 }%
 {}
\usepackage{calc}

\usepackage[top=3.5cm, bottom=3cm, right=1.5cm, left=1.0cm,
            headheight=2.2cm, reversemp, includemp, marginparwidth=4.5cm]{geometry}

\titleformat{\section}
  {\normalfont\sffamily\Large\bfseries}
  {}{0pt}{}
\titleformat{\subsection}
  {\normalfont\sffamily\large\bfseries}
  {}{0pt}{}
\titleformat{\subsubsection}
  {\normalfont\sffamily\bfseries}
  {}{0pt}{}
\titleformat*{\paragraph}
  {\sffamily\normalsize}

\usepackage{fancyhdr}
\makeatletter
\let\ps@plain\ps@fancy
\fancyheadoffset[L]{4.5cm}
\fancyfootoffset[L]{4.5cm}

\definecolor{linky}{rgb}{0.0, 0.5, 1.0}

\newtcolorbox{repobox}
   {colback=red, colframe=red!75!black,
     boxrule=0.5pt, arc=2pt, left=6pt, right=6pt, top=3pt, bottom=3pt}

\newcommand{\ExternalLink}{%
   \tikz[x=1.2ex, y=1.2ex, baseline=-0.05ex]{%
       \begin{scope}[x=1ex, y=1ex]
           \clip (-0.1,-0.1)
               --++ (-0, 1.2)
               --++ (0.6, 0)
               --++ (0, -0.6)
               --++ (0.6, 0)
               --++ (0, -1);
           \path[draw,
               line width = 0.5,
               rounded corners=0.5]
               (0,0) rectangle (1,1);
       \end{scope}
       \path[draw, line width = 0.5] (0.5, 0.5)
           -- (1, 1);
       \path[draw, line width = 0.5] (0.6, 1)
           -- (1, 1) -- (1, 0.6);
       }
   }

\patchcmd{\@maketitle}{center}{flushleft}{}{}
\patchcmd{\@maketitle}{center}{flushleft}{}{}
\patchcmd{\@maketitle}{\LARGE}{\LARGE\sffamily}{}{}
\def\maketitle{{%
  
  \AB@maketitle}}
\makeatletter
\renewcommand\AB@affilsepx{ \protect\Affilfont}
\renewcommand\AB@affilnote[1]{{\bfseries #1}\hspace{3pt}}
\renewcommand{\affil}[2][]%
   {\newaffiltrue\let\AB@blk@and\AB@pand
      \if\relax#1\relax\def\AB@note{\AB@thenote}\else\def\AB@note{#1}%
        \setcounter{Maxaffil}{0}\fi
        \begingroup
        \let\href=\href@Orig
        \let\texttt=\textttOrig
        \let\protect\@unexpandable@protect
        \def\thanks{\protect\thanks}\def\footnote{\protect\footnote}%
        \@temptokena=\expandafter{\AB@authors}%
        {\def\\{\protect\\\protect\Affilfont}\xdef\AB@temp{#2}}%
         \xdef\AB@authors{\the\@temptokena\AB@las\AB@au@str
         \protect\\[\affilsep]\protect\Affilfont\AB@temp}%
         \gdef\AB@las{}\gdef\AB@au@str{}%
        {\def\\{, \ignorespaces}\xdef\AB@temp{#2}}%
        \@temptokena=\expandafter{\AB@affillist}%
        \xdef\AB@affillist{\the\@temptokena \AB@affilsep
          \AB@affilnote{\AB@note}\protect\Affilfont\AB@temp}%
      \endgroup
       \let\AB@affilsep\AB@affilsepx
}
\makeatother

\renewcommand\Affilfont{\sffamily\small\mdseries}
\ifnum 0\ifxetex 1\fi\ifluatex 1\fi=0 
  \usepackage[T1]{fontenc}
  \usepackage[utf8]{inputenc}

\else 
  \ifxetex
    \usepackage{mathspec}
    \usepackage{fontspec}

  \else
    \usepackage{fontspec}
  \fi
  \defaultfontfeatures{Ligatures=TeX,Scale=MatchLowercase}

\fi
\IfFileExists{upquote.sty}{\usepackage{upquote}}{}
\IfFileExists{microtype.sty}{%
\usepackage{microtype}
\UseMicrotypeSet[protrusion]{basicmath} 
}{}

\usepackage{hyperref}
\hypersetup{unicode=true,
            pdftitle={SkyPy: A package for modelling the Universe},
            pdfborder={0 0 0},
            breaklinks=true}
\let\addcontentslineOrig=\addcontentsline
\def\addcontentsline#1#2#3{\bgroup
  \let\texttt=\textttOrig\addcontentslineOrig{#1}{#2}{#3}\egroup}
\let\markbothOrig\markboth
\def\markboth#1#2{\bgroup
  \let\texttt=\textttOrig\markbothOrig{#1}{#2}\egroup}
\let\markrightOrig\markright
\def\markright#1{\bgroup
  \let\texttt=\textttOrig\markrightOrig{#1}\egroup}

\IfFileExists{parskip.sty}{%
\usepackage{parskip}
}{
\setlength{\parindent}{0pt}
\setlength{\parskip}{6pt plus 2pt minus 1pt}
}
\ifx\paragraph\undefined\else
\let\oldparagraph\paragraph
\renewcommand{\paragraph}[1]{\oldparagraph{#1}\mbox{}}
\fi
\ifx\subparagraph\undefined\else
\let\oldsubparagraph\subparagraph
\renewcommand{\subparagraph}[1]{\oldsubparagraph{#1}\mbox{}}
\fi

\title{\texttt{SkyPy}: A package for modelling the Universe}

        \author[1]{Adam Amara\footnote{adam.amara@port.ac.uk}}
          \author[1, 2]{Lucia F. de la Bella}
          \author[3]{Simon Birrer}
          \author[2]{Sarah Bridle}
          \author[2]{Juan Pablo Cordero}
          \author[4]{Ginevra Favole}
          \author[5, 2]{Ian Harrison}
          \author[1]{Ian W. Harry}
          \author[6]{William G. Hartley}
          \author[1]{Coleman Krawczyk}
          \author[1]{Andrew Lundgren}
          \author[7, 8, 9]{Brian Nord}
          \author[1]{Laura K. Nuttall}
          \author[10, 2]{Richard P. Rollins\footnote{richard.rollins@ed.ac.uk}}
          \author[1]{Philipp Sudek}
          \author[11]{Sut-Ieng Tam}
          \author[12]{Nicolas Tessore}
          \author[1]{Arthur E. Tolley}
          \author[11]{Keiichi Umetsu}
          \author[1]{Andrew R. Williamson}
          \author[2]{Laura Wolz}
    
      \affil[1]{Institute of Cosmology and Gravitation, University of
Portsmouth}
      \affil[2]{Jodrell Bank Centre for Astrophysics, University of
Manchester}
      \affil[3]{Kavli Institute for Particle Astrophysics and Cosmology
and Department of Physics, Stanford University}
      \affil[4]{Institute of Physics, Laboratory of Astrophysics, Ecole
Polytechnique Fédérale de Lausanne}
      \affil[5]{Department of Physics, University of Oxford}
      \affil[6]{Department of Astronomy, University of Geneva}
      \affil[7]{Fermi National Accelerator Laboratory}
      \affil[8]{Kavli Institute for Cosmological Physics, University of
Chicago}
      \affil[9]{Department of Astronomy and Astrophysics, University of
Chicago}
      \affil[10]{Institute for Astronomy, University of Edinburgh}
      \affil[11]{Institute of Astronomy and Astrophysics, Academia
Sinica}
      \affil[12]{Department of Physics and Astronomy, University College
London}
  \date{\vspace{-7ex}}

\begin{document}
\maketitle

\marginpar{

  \begin{flushleft}
  \sffamily\small

  {\bfseries DOI:} \href{https://doi.org/10.21105/joss.03056}{\color{linky}{10.21105/joss.03056}}

  \vspace{2mm}

  {\bfseries Software}
  \begin{itemize}
    \setlength\itemsep{0em}
    \item \href{https://github.com/openjournals/joss-reviews/issues/3056}{\color{linky}{Review}} \ExternalLink
    \item \href{https://github.com/skypyproject/skypy}{\color{linky}{Repository}} \ExternalLink
    \item \href{https://doi.org/10.5281/zenodo.4475347}{\color{linky}{Archive}} \ExternalLink
  \end{itemize}

  \vspace{2mm}

  \par\noindent\hrulefill\par

  \vspace{2mm}

  {\bfseries Editor:} \href{http://www.arfon.org}{Arfon Smith} \ExternalLink \\
  \vspace{1mm}
    {\bfseries Reviewers:}
  \begin{itemize}
  \setlength\itemsep{0em}
    \item \href{https://github.com/cescalara}{@cescalara}
    \item \href{https://github.com/rmorgan10}{@rmorgan10}
    \end{itemize}
    \vspace{2mm}

  {\bfseries Submitted:} 15 February 2021\\
  {\bfseries Published:} 10 September 2021

  \vspace{2mm}
  {\bfseries License}\\
  Authors of papers retain copyright and release the work under a Creative Commons Attribution 4.0 International License (\href{http://creativecommons.org/licenses/by/4.0/}{\color{linky}{CC BY 4.0}}).

  \end{flushleft}
}

\hypertarget{summary}{%
\section{Summary}\label{summary}}

\texttt{SkyPy} is an open-source Python package for simulating the
astrophysical sky. It comprises a library of physical and empirical
models across a range of observables and a command line script to run
end-to-end simulations. The library provides functions that sample
realisations of sources and their associated properties from probability
distributions. Simulation pipelines are constructed from these models
using a YAML-based configuration syntax, while task scheduling and data
dependencies are handled internally and the modular design allows users
to interface with external software. \texttt{SkyPy} is developed and
maintained by a diverse community of domain experts with a focus on
software sustainability and interoperability. By fostering
co-development, it provides a framework for correlated simulations of a
range of cosmological probes including galaxy populations, large scale
structure, the cosmic microwave background, supernovae and gravitational
waves.

Version \texttt{0.4} implements functions that model various properties
of galaxies including luminosity functions, redshift distributions and
optical photometry from spectral energy distribution templates. Future
releases will provide additional modules, for example to simulate
populations of dark matter halos and model the galaxy-halo connection,
making use of existing software packages from the astrophysics community
where appropriate.

\hypertarget{statement-of-need}{%
\section{Statement of need}\label{statement-of-need}}

An open-data revolution in astronomy led by past, ongoing, and future
legacy surveys such as \emph{Euclid} (Laureijs et al., 2011), the Rubin
Observatory Legacy Survey of Space and Time (Ivezić et al., 2019),
\emph{Planck} (Planck Collaboration, 2020) and the Laser Interferometer
Gravitational-Wave Observatory (LIGO Scientific Collaboration, 2015)
means access to data is no longer the primary barrier to research.
Instead, access to increasingly sophisticated analysis methods is
becoming a significant challenge. Researchers frequently need to model
multiple astronomical probes and systematics to perform a statistically
rigorous analysis that fully exploits the available data. In particular,
forward modelling and machine learning have emerged as important
techniques for the next generation of surveys and both depend on
realistic simulations. However, existing software is frequently
closed-source, outdated, unmaintained or developed for specific projects
and surveys making it unsuitable for the wider research community. As a
consequence astronomers routinely expend significant effort replicating
or re-developing existing code. The growing need for skill development
and knowledge sharing in astronomy is evidenced by a number of open
initiatives focused on software, statistics and machine learning e.g.,
Astropy (Astropy Collaboration, 2018, 2013), OpenAstronomy
(https://openastronomy.org), Dark Machines (http://darkmachines.org),
The Deep Skies Lab (https://deepskieslab.com), and the Cosmo-Statistics
Initiative (https://cosmostatistics-initiative.org). Recently, the
research community has developed a number of important open-source
python packages that address individual aspects of modelling
astronomical surveys e.g.~\texttt{GalSim} (Rowe et al., 2015),
\texttt{Halotools} (Hearin et al., 2017) and \texttt{popsynth} (Burgess
\& Capel, 2021). \texttt{SkyPy} was established as a part of this
software ecosystem to meet the need for realistic end-to-end simulations
and enable forward modelling and machine learning applications.

\hypertarget{acknowledgements}{%
\section{Acknowledgements}\label{acknowledgements}}

AA and PS acknowledge support from a Royal Society Wolfson Fellowship
grant. LFB, SB, IH, and RPR acknowledge support from the European
Research Council in the form of a Consolidator Grant with number 681431.
IH also acknowledges support from the Beecroft Trust. JPC acknowledges
support granted by Agencia Nacional de Investigación y Desarrollo (ANID)
DOCTORADO BECAS CHILE/2016 - 72170279. GF acknowledges financial support
from the SNF 175751 ``Cosmology with 3D maps of the Universe'' research
grant. AL and ARW thanks the STFC for support through the grant
ST/S000550/1. LKN thanks the UKRI Future Leaders Fellowship for support
through the grant MR/T01881X/1. This manuscript has been authored by
Fermi Research Alliance, LLC under Contract No.~DE-AC02-07CH11359 with
the U.S. Department of Energy, Office of Science, Office of High Energy
Physics.

\hypertarget{references}{%
\section*{References}\label{references}}
\addcontentsline{toc}{section}{References}

\hypertarget{refs}{}
\begin{CSLReferences}{1}{0}
\leavevmode\hypertarget{ref-Astropy2018}{}%
Astropy Collaboration. (2018). {The Astropy Project: Building an
Open-science Project and Status of the v2.0 Core Package}. \emph{The
Astronomical Journal}, \emph{156}.
\url{https://doi.org/10.3847/1538-3881/aabc4f}

\leavevmode\hypertarget{ref-Astropy2013}{}%
Astropy Collaboration. (2013). {Astropy: A community Python package for
astronomy}. \emph{Astronomy and Astrophysics}, \emph{558}.
\url{https://doi.org/10.1051/0004-6361/201322068}

\leavevmode\hypertarget{ref-popsynth2021}{}%
Burgess, J., \& Capel, F. (2021). {popsynth: A generic astrophysical
population synthesis framework}. \emph{The Journal of Open Source
Software}, \emph{6}(63), 3257. \url{https://doi.org/10.21105/joss.03257}

\leavevmode\hypertarget{ref-halotools2017}{}%
Hearin, A. P., Campbell, D., Tollerud, E., Behroozi, P., Diemer, B.,
Goldbaum, N. J., Jennings, E., Leauthaud, A., Mao, Y.-Y., More, S.,
Parejko, J., Sinha, M., Sipöcz, B., \& Zentner, A. (2017). {Forward
Modeling of Large-scale Structure: An Open-source Approach with
Halotools}. \emph{The Astronomical Journal}, \emph{154}(5), 190.
\url{https://doi.org/10.3847/1538-3881/aa859f}

\leavevmode\hypertarget{ref-LSST2019}{}%
Ivezić, Ž., Kahn, S. M., Tyson, J. A., Abel, B., Acosta, E., Allsman,
R., Alonso, D., AlSayyad, Y., Anderson, S. F., Andrew, J., Angel, J. R.
P., Angeli, G. Z., Ansari, R., Antilogus, P., Araujo, C., Armstrong, R.,
Arndt, K. T., Astier, P., Aubourg, É., \ldots{} Zhan, H. (2019). {LSST:
From Science Drivers to Reference Design and Anticipated Data Products}.
\emph{The Astrophysical Journal}, \emph{873}, 111.
\url{https://doi.org/10.3847/1538-4357/ab042c}

\leavevmode\hypertarget{ref-Euclid2011}{}%
Laureijs, R., Amiaux, J., Arduini, S., Auguères, J.-L., Brinchmann, J.,
Cole, R., Cropper, M., Dabin, C., Duvet, L., Ealet, A., Garilli, B.,
Gondoin, P., Guzzo, L., Hoar, J., Hoekstra, H., Holmes, R., Kitching,
T., Maciaszek, T., Mellier, Y., \ldots{} Zucca, E. (2011). {Euclid
Definition Study Report}. \emph{arXiv e-Prints}, arXiv:1110.3193.
\url{http://arxiv.org/abs/1110.3193}

\leavevmode\hypertarget{ref-LIGO2015}{}%
LIGO Scientific Collaboration. (2015). {Advanced LIGO}. \emph{Classical
and Quantum Gravity}, \emph{32}(7), 074001.
\url{https://doi.org/10.1088/0264-9381/32/7/074001}

\leavevmode\hypertarget{ref-Planck2020}{}%
Planck Collaboration. (2020). {Planck 2018 results. I. Overview and the
cosmological legacy of Planck}. \emph{Astronomy \& Astrophysics},
\emph{641}, A1. \url{https://doi.org/10.1051/0004-6361/201833880}

\leavevmode\hypertarget{ref-Galsim2015}{}%
Rowe, B. T. P., Jarvis, M., Mandelbaum, R., Bernstein, G. M., Bosch, J.,
Simet, M., Meyers, J. E., Kacprzak, T., Nakajima, R., Zuntz, J.,
Miyatake, H., Dietrich, J. P., Armstrong, R., Melchior, P., \& Gill, M.
S. S. (2015). {GALSIM: The modular galaxy image simulation toolkit}.
\emph{Astronomy and Computing}, \emph{10}, 121--150.
\url{https://doi.org/10.1016/j.ascom.2015.02.002}

\end{CSLReferences}

\end{document}